# Chaotic Excitations of Rogue Waves in Stable Parametric Region for Highly-Energetic Pair Plasmas

S. Dey [1], D. Maity [2], A. Ghosh [3], P. Samanta [4], A. De [5] and S. Chandra [6*]

1,2 Department of Physics, Visva-Bharati University, Santiniketan, Bolpur, West Bengal-731235, India
3 Department of Physics, Bethune College University of Calcutta, Manicktala, Kolkata, 700006, India
4,5 Physics Department, Scottish Church college, Kolkata 700006, India
6 Department of Physics, Government General Degree College at Kushmandi, India, 733121.
*Institute of Natural Sciences and Applied Technology, Kolkata, India, 700032.

E-mail: ssuman7226@gmail.com (corresponding author)

We have studied the Rogue wave existence and propagation in Ion-acoustic (IA) mode for the highly energetic case using kappa distributed electrons in accordance with the Korteweg–de Vries (KdV) equation that is modified KdV and extended KdV equation. We have used reductive perturbation method. We first examined the linear dispersive behaviour in Ion-acoustic mode. Obtaining the Nonlinear Schrödinger equation, we simulated Rogue wave and examined dynamics of it, and its response to small perturbations. We discussed the possibility of generation of Rogue wave as well as the stability of this against various parameters like wave number, spatial and time component. This study is quite helpful for understanding some prominent points of the nonlinearity of IA waves and Rogue wave generation of the highly energetic case in space plasma also in a laboratory plasma.

## 1. Introduction

We have studied the possibilities of the ion-acoustic rogue waves (RWs) propagation under the distributions with high-energy tails which is the generally obtained Lorentzian (or Kappa) distribution functions, frequently the κ lies in the range of $1.5 < \kappa < 6$. Zaheer et al [1] used Kappa-distribution functions to study electrostatic modes such as Langmuir waves, dust ion-acoustic waves, and dust acoustic waves in his paper. For $\kappa \to \infty$ it will drop to Maxwellian distribution, a Kappa $(\kappa)$ distribution function to describe super thermal populations. The family of kappa velocity distributions, introduced first by Vasyliunas (1968), is recognized to be highly appropriate for modelling specific electron and ion components of different plasma states. Now a Kappa $(\kappa)$ distribution which provides an unquestionable replacement for a Maxwellian distribution in space plasmas provides a continuous spectrum of energy.

"Rogue wave" (RW) is the large amplitude wave that occur more frequently than expected for normal. RWs have a peak value of amplitude more than twice of the significant wave height. It appears from nowhere and disappear without a trace [2-4] and has a localization property.

There is compressive and rarefactive solitary wave structure is possible in plasmas with immobile ions and streaming electrons [5]. To study the non-linear characteristics, we first study the linear dispersive characteristics in Ion-Acoustic mode and the variation of wave frequency $(\omega)$ with wave number (k) with the dependence of other

$$p_j = \frac{m_j V_{Fj}^2}{3 n_{j0}^2} n_j^2 \qquad (1)$$

unperturbed quantities. For electron plasma waves in a quantum plasma the variation of dispersion properties with the quantum diffraction is discussed by S. Chandra [6].

Korteweg–de Vries (KdV)/modified and Extended Korteweg–de Vries equation and non-linear Schrödinger equation (NLSE) has been used to interpret the results of non-linear systems such as Ocean, Space and Astrophysical plasmas, Solar winds. The dynamics of non-linear RWs is governed by NLSE. Using appropriate transformations on NLSE we arrived at a set of ordinary differential equations which were solved to obtain the phase trajectory and time series. It provides interesting results about the chaotic nature of the system. It is found that the gravity field and external magnetic field have a significant effect on the amplitude of the rogue waves. The numerical results obtained in the paper [7]. Electron-positron *(e-p)* plasma has its unavoidable interest for its wide-ranging applications in many astrophysical contexts, such as the early universe [8], active galactic nuclei structures [9], pulsar magnetospheres [10] and many more. In understanding of the nonlinear Langmuir rogue waves which accompanies collision less electron-positron *(e-p)* plasmas are presented in moslem2011 [11] paper in detail.

The aim of our present work is to understand the structure of the Langmuir RWs and excitation of non-linear RWs in a plasma strongly composed of electrons and positrons. Particular questions to be answered are: *(i)* non-linear structures and density perturbations in an *(e-p)* as and *(ii)* nature of wave envelops of non-linear RWs and further importance of the *(e-p)* plasma applications in astrophysical environments in general.

## 2. Basic Formulations

We have considered here non-linear ion-acoustic waves propagating in unmagnetized plasma have no viscous effect at finite temperature also electrons are kappa distributed. Now the particles are considered as one-dimensional Fermi gas model at zero temperature with the pressure term [12], where the index $j = e$ for electron and $j = i$ for ions, $m_j = m_i$ is the mass of electrons and $m_j = m_i$ is the mass of ion, $T_{Fj}$ is the Fermi temperature, $V_{Fj}$ is the Fermi speed. Number density of ions and electrons are respectively $n_i$ and $n_e$.

For Ion-acoustic mode the non-normalized basic governing equations are given by as follows,

$$\frac{\partial n_i}{\partial t} + \frac{\partial (n_i u_i)}{\partial x} = 0 \qquad (2)$$

$$\left(\frac{\partial}{\partial t} + u_i \frac{\partial}{\partial x}\right) u_i = \frac{Q_i}{m_i} \frac{\partial \phi}{\partial x} \qquad (3)$$

$$\frac{\partial^2 \phi}{\partial x^2} = 4\pi \left(Q_e n_e + Q_i n_i\right) \qquad (4)$$

Here $\phi$ is the electrostatic wave potential, for electron charge can be written as $Q_e = e$ and for ion charge will be $Q_i = -Z_i e$ and $u_i, n_i, p_i$ are the fluid velocity, number density and pressure terms for ions. For convenience we introduce the normalisation schemes as follows:

$$\bar{x} \to \frac{x\omega_i}{V_{Fh}}, \bar{t} \to t\omega_i, \bar{\phi} \to \frac{e\phi}{2k_B T_{Feh}},$$

$$\bar{n}_i \to \frac{n_i}{n_{i0}}, \bar{n}_j \to \frac{n_j}{n_{j0}}, \bar{u}_j \to \frac{u_j}{V_{Fh}}$$

In which $\omega_{ec} = \sqrt{\frac{4\pi n_{ec0} e^2}{m_e}}$ is the Plasma frequency, $V_{Fh} = \sqrt{\frac{2k_B T_{Feh}}{m_e}}$ is the quantum ion acoustic velocity, $T_{Feh}$ is the Fermi temperature of electron and $n_{io}, n_{eo}$ are the equilibrium density of ion and electrons respectively.

Now we can obtain the equations in a dimensionless form as follows:

$$\frac{\partial \bar{n}_i}{\partial \bar{t}} + \frac{\partial (\bar{n}_i \bar{u}_i)}{\partial \bar{x}} = 0 \qquad (5)$$

$$\left(\frac{\partial}{\partial \bar{t}} + \bar{u}_i \frac{\partial}{\partial \bar{x}}\right) \bar{u}_i = -\mu \frac{\partial \bar{\phi}}{\partial \bar{x}} \qquad (6)$$

$$\frac{\partial^2 \bar{\phi}}{\partial \bar{x}^2} = (\bar{n}_e - \bar{n}_i) \qquad (7)$$

where $\mu = \frac{m_e}{m_i}$

Highly energetic electrons are in plasmas can be treated with Kappa $(\kappa)$ distribution which is formed in isotropic case,

$$f_e^\kappa (r,v) = \frac{n_e}{2\pi (\kappa w_{Ke}^2)^{3/2}} \frac{\Gamma(\kappa+1)}{\Gamma\left(\kappa - \frac{1}{2}\right)\Gamma\left(\frac{3}{2}\right)} \left(1 + \frac{v_e^2}{\kappa w_{Ke}^2}\right)^{-(\kappa+1)} \qquad (8)$$

Where, $w_{Ke} = \sqrt{(2\kappa-3)kT_e/\kappa m_e}$ is the thermal velocity, $m_e$ is the mass of electrons, $T_e$ is the equivalent temperature of electron, $v_e$ is the velocity of electrons, $\Gamma(\kappa)$ is the Gamma function.

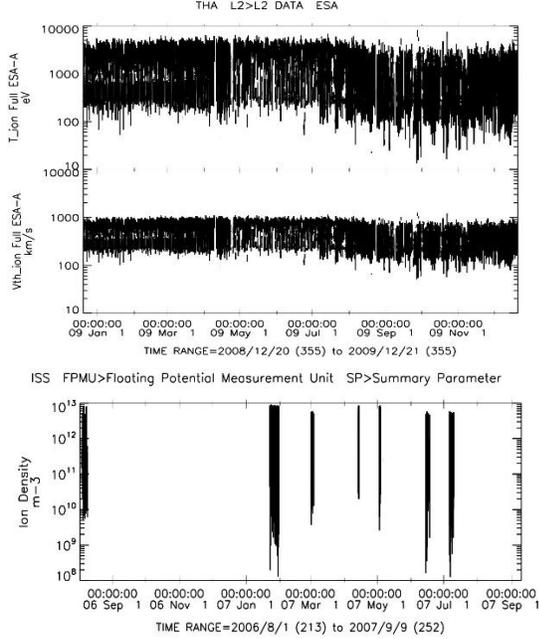

**Fig.1:** Ranges for Thermal Ion-Velocity, $T_{ion}$, Ion density for Solar wind plasma. Data Source: *THEMIS Themis Probe A, Type: ESA, PI: C.W. Carlson & J. McFadden, PI_AFFILIATION: NASA NAS5-02099.*

As it is clear that for a well-defined solution, we need the value of $\kappa > 3/2$ at that value the distribution function collapses and the equivalent temperature is not defined. See at $\kappa \to \infty$ the distribution function is reduced to Maxwellian distribution form. If $\sigma_e = T_{Feh}/T_e$ then from this the normalization electron density can be obtained as,

$$\bar{n}_{ke} = \left(1 - \frac{\bar{\phi}\sigma_e}{\kappa - 3/2}\right)^{-\kappa + 1/2} \quad (9)$$

$$\left(1 - \frac{\bar{\phi}\sigma_e}{\kappa - 3/2}\right)^{-\kappa + 1/2} \to e^{+\bar{\phi}} \quad (10)$$

That means as $\kappa \to \infty$ equation to can be degenerated into a Maxwellian distribution. Now, $\kappa > 3/2$ is needed for a physically valid solution. Further we expand equation (9) in binomial expansion we get,

$$\bar{n}_{ke} = 1 + \left(\kappa - \frac{1}{2}\right)\frac{\bar{\phi}\sigma_e}{(\kappa - 3/2)}$$
$$+ \frac{1}{2}\frac{(\kappa - 1/2)(\kappa + 1/2)}{(\kappa - 3/2)^2}\bar{\phi}^2\sigma_e^2 + \ldots \quad (11)$$

$$\bar{n}_{ke} = 1 + A(\kappa)\bar{\phi}\sigma_e + \frac{1}{2}B(\kappa)\bar{\phi}^2\sigma_e^2 + \ldots \quad (12)$$

Where $A(\kappa) = \dfrac{\left(\kappa - \dfrac{1}{2}\right)}{(\kappa - 3/2)}$ and

$$B(\kappa) = \frac{(\kappa - 1/2)(\kappa + 1/2)}{(\kappa - 3/2)^2}$$

and for lower order terms we can obtain the solution as,

$$\bar{n}_{ke} = 1 + \left(\kappa - \frac{1}{2}\right)\frac{\bar{\phi}\sigma_e}{(\kappa - 3/2)} \quad (13)$$

### 3. Analytical Study

#### 3.1 Linear dispersion relation

Considering that linear characteristics is only for electrons and as ions enters into the region it will be treated as non-linear way. To obtain linear dispersive characteristics of electron as well as non-linearity of respective ions we have considered a perturbation expression for $\bar{n}_{\kappa e}, \bar{n}_i, \bar{u}_i, \bar{\phi}$ quantities,

$$\begin{bmatrix}\bar{n}_{\kappa e}\\ \bar{n}_i\\ \bar{u}_i\\ \bar{\phi}\end{bmatrix} = \begin{bmatrix}1\\ 1\\ u_0\\ \phi_0\end{bmatrix} + \begin{bmatrix}\bar{n}_e^{(1)}\\ \bar{n}_i^{(1)}\\ \bar{u}_i^{(1)}\\ \bar{\phi}^{(1)}\end{bmatrix} + \begin{bmatrix}\bar{n}_e^{(2)}\\ \bar{n}_i^{(2)}\\ \bar{u}_i^{(2)}\\ \bar{\phi}^{(2)}\end{bmatrix} + \ldots \quad (14)$$

Substituting the expression (14) for each term in equations (5)-(7) and linearizing and considering that,

$$\bar{n}_{\kappa e} = 1 + \grave{o}\bar{n}_e^{(1)}$$
$$\bar{n}_i = 1 + \grave{o}\bar{n}_i^{(1)}$$
$$\bar{u}_i = u_0 + \grave{o}\bar{u}_1^{(1)}$$
$$\bar{\phi} = \phi_0 + \grave{o}\bar{\phi}_1^{(1)}$$

Also, we assume that all field quantities vary as following manner,

$$\bar{n}_e^{(1)} = \bar{n}_e^{(1)} e^{(k\bar{x}-\omega\bar{t})}$$
$$\bar{n}_i^{(1)} = \bar{n}_i^{(1)} e^{(k\bar{x}-\omega\bar{t})}$$
$$\bar{u}_i^{(1)} = \bar{u}_i^{(1)} e^{(k\bar{x}-\omega\bar{t})}$$
$$\bar{\phi}^{(1)} = \bar{\phi}^{(1)} e^{(k\bar{x}-\omega\bar{t})}$$

where we get the normalised wave frequency ω and wave number is k. Further we can replace the quantities $\frac{\partial}{\partial \bar{x}}$ by its corresponding eigenvalue $(ik)$ and $\frac{\partial}{\partial \bar{t}}$ by its eigenvalue $(-i\omega)$ by replacing these quantities we get,

$$\bar{u}_i^{(1)} = \frac{k\mu}{(\omega - ku_0)} \bar{\phi}^{(1)} \quad (15)$$

And

$$\bar{n}_i^{(1)} = \frac{k^2}{(\omega - ku_0)^2} \bar{\phi}^{(1)} \quad (16)$$

and further putting into normalised poisson equation (7) we obtain following dispersion relation:

$$\omega = u_0 k \pm \sqrt{\frac{\mu k^2}{k^2 + \left(\frac{\kappa - 1/2}{\kappa - 3/2}\right)\sigma_e}} \quad (17)$$

which is the linear dispersion relation for ion-acoustic wave in high energetic case. So, equation (17) indicates that there will be two stable liner dispersions for ion-acoustic wave, where ions provide the inertia. Here we used the kappa distribution for electron for first order perturbative expansion for that we had taken the solution (13) as, considering the first order perturbation expansion and comparing both side we get,

$$\bar{n}_{ke}^{(1)} = \frac{\left(\kappa - \frac{1}{2}\right)}{(\kappa - 3/2)} \bar{\phi}^{(1)} \sigma_e \quad (18)$$

which we have used in normalised poisson equation (7) to obtain the following dispersion relation.
The phase speed of the wave is:

$$V_p = \frac{\omega}{k} = \frac{u_0 k \pm \sqrt{\frac{\mu k^2}{k^2 + \left(\frac{\kappa - 1/2}{\kappa - 3/2}\right)\sigma_e}}}{k}$$

Here we assumed the electron to be non-inertial and ions are inertial. For that $V_p$ should lie between $V_{Fi} \ll V_p \ll V_{Fe}$. $V_{Fi}$ and $V_{Fe}$ are the Fermi velocities for ions and electrons respectively.

Later we examined the dispersive nature by plotting $\omega - k$ curve for different $\kappa$ values.

### 3.2 Derivation of KdV equation

To examine the non-linearity behaviour of ion-acoustic wave we have considered inertia less hot non-relativistic electrons also ion mass is considered as positron mass as similar to an electron mass, there will be similar treatment in [11], [13].

Let assume $\bar{n}_e, \bar{n}_i, \bar{u}_i, \bar{\phi}$ are slowly variable respect to $\bar{x}$ and $\bar{t}$ so that they can be stretched by the variables called stretching variables as,

$$\xi = \grave{o}^{1/2}(\bar{x} - V_0 \bar{t}), \; \tau = \grave{o}^{3/2}\bar{t} \quad (19)$$

Where, ò is the dimensionless smallness parameter and $V_0$ is the normalised linear velocity. Now writing equations (5)-(7) in terms of these stretching variables and solving for the lowest $\epsilon$ order and apply boundary conditions for first order field variables $\bar{n}_e^{(1)}, \bar{n}_i^{(1)}, \bar{u}_i^{(1)}, \bar{\phi}^{(1)}$ and $|\xi| \to \infty$.

Now Poisson equation gives,

$$\frac{\partial^3 \bar{\phi}^{(1)}}{\partial \xi^3} = \frac{\sigma_e^2}{2} B(\kappa) \bar{\phi}^{(1)} \frac{\partial \bar{\phi}^{(1)}}{\partial \xi} - \frac{\partial \bar{n}_i^{(2)}}{\partial \xi} \quad (20)$$

Also, we get these equations by comparing higher order terms of ò
From continuity equation,

$$\delta^{3/2}\left[-V_0\frac{\partial \bar{n}_i^{(1)}}{\partial \xi}+\frac{\partial}{\partial \xi}\left(u_0\bar{n}_i^{(1)}+\bar{u}_i^{(1)}\right)\right]=0 \quad (21)$$

$$\delta^{5/2}\left[\frac{\partial \bar{n}_i^{(1)}}{\partial \tau}-V_0\frac{\partial \bar{n}_i^{(2)}}{\partial \xi}+\frac{\partial}{\partial \xi}\left(\bar{n}_i^{(1)}\bar{u}_i^{(1)}+u_0\bar{n}_i^{(2)}+\bar{u}_i^{(2)}\right)\right]=0 \quad (22)$$

From momentum equation,

$$\delta^{3/2}\left[-V_0\frac{\partial \bar{u}_i^{(1)}}{\partial \xi}+u_0\frac{\partial \bar{u}_i^{(1)}}{\partial \xi}+\mu\frac{\partial \bar{\phi}^{(1)}}{\partial \xi}\right]=0 \quad (23)$$

$$\delta^{5/2}\left[\frac{\partial \bar{u}_i^{(1)}}{\partial \tau}-V_0\frac{\partial \bar{u}_i^{(2)}}{\partial \xi}+\bar{u}_i^{(1)}\frac{\partial \bar{u}_i^{(1)}}{\partial \xi}+u_0\frac{\partial \bar{u}_i^{(2)}}{\partial \xi}+\mu\frac{\partial \bar{\phi}^{(2)}}{\partial \xi}\right]=0 \quad (24)$$

We will get from equation (22) and (24),

$$\bar{u}_i^{(1)}=\frac{\mu}{(V_0-u_0)}\bar{\phi}^{(1)} \quad (25)$$

And

$$\bar{n}_i^{(1)}=\frac{\mu}{(V_0-u_0)^2}\bar{\phi}^{(1)} \quad (26)$$

Using the reductive perturbation method, comparing suitable powers of $\epsilon$ and doing algebraic operations and rearranging equations (21)-(26) we will obtain the KdV equation for ion-acoustic wave,

$$\frac{\partial \psi}{\partial \tau}+N\psi\frac{\partial \psi}{\partial \xi}+D\frac{\partial^3 \psi}{\partial \xi^3}=0 \quad (27)$$

where we used $\bar{\phi}^{(1)}=\psi$.

$$N=\left(\frac{3\mu}{2(V_0-u_0)}-\frac{\sigma_e(V_0-u_0)^3 B(\kappa)}{2\mu}\right) \quad (28)$$

$$D=\frac{(V_0-u_0)^3}{2\mu} \quad (29)$$

Where, $N$ is the nonlinear term and $D$ is the dispersive term and $B(\kappa)=\frac{(\kappa-1/2)(\kappa+1/2)}{(\kappa-3/2)^2}$ have been used.

To obtain the steady state solution of KdV equation we now introduce a transformation $\chi=\xi-M\tau$ where $M$ is the normalised speed of wave, it is called the Mach number. It can be expressed as $v/V_{Fi}$, where $v$ is the non-normalized wave velocity. These two terms are responsible for dispersive and nonlinear effect in ion acoustic mode. Again, applying boundary conditions $|\chi|\to\pm\infty$, $\psi$, $\frac{\partial \psi}{\partial \chi}$, $\frac{\partial^2 \psi}{\partial \chi^2}\to 0$ to obtain a possible stationary solution [14],

$$\psi=\psi_m sech^2\frac{\chi}{\Delta} \quad (30)$$

Where, $\psi_m=\frac{3M}{N}$ and $\Delta=\sqrt{\frac{4D}{M}}$

Now, as we saw that at a critical value of $\kappa=\kappa_c$ the Nonlinear term of KdV equation is not acceptable so KdV equation is invalid under that critical value of $\kappa$. So, we have to modify it like the critical condition is independent of Nonlinear term.

### 3.3 Extended KdV equation

To derive the extended KdV equation we have taken the stretching variable as:

$$\xi=\delta(\bar{x}-V_0\bar{t}), \quad \tau=\delta^3\bar{t} \quad (31)$$

As previous we have equated the $\epsilon$ terms from lower order to higher order and we got those equations:

From continuity equation,

$$\delta^2\left[-V_0\frac{\partial \bar{n}_i^{(1)}}{\partial \xi}+\frac{\partial}{\partial \xi}\left(u_0\bar{n}_i^{(1)}+\bar{u}_i^{(1)}\right)\right]=0 \quad (32)$$

$$\delta^3\left[-V_0\frac{\partial \bar{n}_i^{(2)}}{\partial \xi}+\frac{\partial}{\partial \xi}\left(\bar{n}_i^{(1)}\bar{u}_i^{(1)}+u_0\bar{n}_i^{(2)}+\bar{u}_i^{(2)}\right)\right]=0 \quad (33)$$

$$\delta^4\left[-(V_0-u_0)\frac{\partial \bar{n}_i^{(3)}}{\partial \xi}+\frac{\partial \bar{n}_i^{(1)}}{\partial \tau}-\frac{\partial}{\partial \xi}\left(\bar{n}_i^{(1)}\bar{u}_i^{(2)}+\bar{u}_i^{(1)}\bar{n}_i^{(2)}\right)+\bar{u}_i^{(2)}+\frac{\partial \bar{u}_i^{(3)}}{\partial \xi}\right]=0 \quad (34)$$

From momentum equation,

$$\delta^2\left[-V_0\frac{\partial \bar{u}_i^{(1)}}{\partial \xi}+u_0\frac{\partial \bar{u}_i^{(1)}}{\partial \xi}+\mu\frac{\partial \bar{\phi}^{(1)}}{\partial \xi}\right]=0 \quad (35)$$

Similarly,

$$\delta^3\left[-V_0\frac{\partial \bar{u}_i^{(2)}}{\partial \xi}+\bar{u}_i^{(1)}\frac{\partial \bar{u}_i^{(1)}}{\partial \xi}+u_0\frac{\partial \bar{u}_i^{(2)}}{\partial \xi}+\mu\frac{\partial \bar{\phi}^{(2)}}{\partial \xi}\right]=0 \quad (36)$$

$$\delta^4\left[-(V_0-u_0)\frac{\partial}{\partial \xi}\left(\bar{u}_i^{(3)}+\bar{u}_i^{3(1)}\right)+\frac{\partial \bar{u}_i^{(1)}}{\partial \tau}+\frac{\partial}{\partial \tau}\left(\bar{u}_i^{(1)}\bar{u}_i^{(2)}\right)+\frac{\partial}{\partial \tau}\left(\bar{u}_i^{3(1)}\right)+\frac{\partial \bar{\phi}^{(3)}}{\partial \xi}\right]=0 \quad (37)$$

From equation (33), (34), (36), (37) we will get,

$$\bar{u}_i^{(1)} = \frac{\mu}{(V_0 - u_0)} \bar{\phi}^{(1)} \quad (38)$$

$$\bar{n}_i^{(1)} = \frac{\mu}{(V_0 - u_0)^2} \bar{\phi}^{(1)} \quad (39)$$

$$\bar{u}_i^{(2)} = \frac{\mu^2}{2(V_0 - u_0)^3} \bar{\phi}^{2(1)} + \frac{\mu}{(V_0 - u_0)} \bar{\phi}^{(2)} \quad (40)$$

$$\bar{n}_i^{(2)} = \frac{3\mu^2}{2(V_0 - u_0)^4} \bar{\phi}^{2(1)} + \frac{\mu}{(V_0 - u_0)^2} \bar{\phi}^{(2)} \quad (41)$$

Now let $R = (V_0 - u_0)$, so from KdV equation the critical value for which the nonlinear term vanishes that will be $R^4 = \frac{3\mu^2}{\sigma_e^2 B(\kappa)}$.

The poisson equation will be written as for $\grave{o}^3$,

$$\frac{\partial^2 \bar{\phi}^{(1)}}{\partial \bar{x}^2} = A(\kappa)\sigma_e^{(3)}\bar{\phi} + \sigma_e^2 B(\kappa)\bar{\phi}^{(1)}\bar{\phi}^{(2)} - \bar{n}_i^{(3)}$$

$$\Rightarrow \frac{\partial^3 \bar{\phi}^{(1)}}{\partial \bar{x}^3} = A(\kappa)\sigma_e \frac{\partial \bar{\phi}^{(3)}}{\partial \bar{x}} + \sigma_e^2 B(\kappa) \frac{\partial}{\partial \bar{x}}(\bar{\phi}^{(1)}\bar{\phi}^{(2)}) - \bar{n}_i^{(3)}$$

Now using equations (33)-(41) and using the fact that the coefficient of $\partial \bar{\phi}^{(3)}/\partial \xi$ due to the critical condition as well as the coefficient of $\partial(\bar{\phi}^{(2)}\bar{\phi}^{(1)})/\partial \xi$ also vanishes. So, we will finally be reached to the modified KdV equation,

$$\frac{\partial \psi}{\partial \tau} + N_1 \psi^2 \frac{\partial \psi}{\partial \xi} + D_1 \frac{\partial^3 \psi}{\partial \xi^3} = 0 \quad (42)$$

where we used $\bar{\phi}^{(1)} = \psi$. Here,

$$N_1 = \frac{R^2}{2\mu}\left(\frac{\mu^3}{(V_0 - u_0)^3} + \frac{3\mu^3}{2(V_0 - u_0)^5} - \frac{\mu^3}{2(V_0 - u_0)^4}\right) \quad (43)$$

$$D_1 = \frac{(V_0 - u_0)^3}{2\mu} \quad (44)$$

Now we take $\delta$ as the relative charge density and rewriting the poisson equation as,

$$\frac{\partial^2 \bar{\phi}}{\partial \bar{x}^2} = -\delta = (\bar{n}_e - \bar{n}_i) \quad (45)$$

also adapting perturbation as, $\delta = \grave{o}\delta^{(1)} + \grave{o}^2 \delta^{(2)} + \ldots$
At the critical value $\kappa = \kappa_c$ we equate $\delta$ for $\grave{o}^2$ we get,

$$\delta^{(2)} = -\frac{1}{2} N \{\bar{\phi}^{(2)}\}^2 = 0 \quad (46)$$

$$N_0 \square s \left(\frac{\partial N}{\partial \kappa}\right)|\kappa - \kappa_c| = c_1 s \grave{o} \quad (47)$$

$|\kappa - \kappa_c|$ is here small dimensionless parameter, $|\kappa - \kappa_c| \square \grave{o}$ and $s = 1$ for $\kappa > \kappa_c$ and $s = -1$ for $\kappa < \kappa_c$ so,

$$\grave{o}\delta^{(2)} \square -\grave{o}^{3/2} \frac{1}{2} c_1 s \{\bar{\phi}^{(2)}\}^2 \quad (48)$$

$$\frac{\partial^2 \bar{\phi}^{(1)}}{\partial \bar{x}^2} + \frac{1}{2} c_1 s \{\bar{\phi}^{(2)}\}^2 - \bar{n}_{\kappa_e}^{(3)} - \bar{n}_i^{(3)} = 0$$

after rearranging terms, we get the extended KdV equation as,

$$\frac{\partial \bar{\phi}^{(1)}}{\partial \tau} + c_1 s N \bar{\phi}^{(1)} \frac{\partial \bar{\phi}^{(1)}}{\partial \xi} + N_1 \{\bar{\phi}^{(1)}\}^2 \frac{\partial \bar{\phi}^{(1)}}{\partial \xi} + D_1 \frac{\partial^3 \psi}{\partial \xi^3} = 0$$

$$\Rightarrow \frac{\partial \bar{\phi}^{(1)}}{\partial \tau} + C_{1\bar{\phi}}^{(1)} \frac{\partial \bar{\phi}^{(1)}}{\partial \xi} + N_1 \{\bar{\phi}^{(1)}\}^2 \frac{\partial \bar{\phi}^{(1)}}{\partial \xi} + D_1 \frac{\partial^3 \psi}{\partial \xi^3} = 0$$

Where, we have taken, $C_1 = c_1 s N$ and $\bar{\phi}^{(1)} = \psi$.

$$\frac{\partial \psi}{\partial \tau} + (C_1 \psi + N_1 \psi^2) \frac{\partial \psi}{\partial \xi} + D_1 \frac{\partial^3 \psi}{\partial \xi^3} = 0 \quad (49)$$

This is the extended KdV equation now if $s = 1$ and $N_1 = 0$ it will be reduced to KdV equation and if we take $C_1 = 0$ it will act as modified KdV equation. The condition of criticality is now removed and we can go for instable and stable condition both.

### 3.4 NLSE and 'Peregrine' Soliton of Rogue Wave

Rogue waves occurred in abundance in the plasma surface. One way to study Rogue waves mathematically is based on nonlinear Schrodinger equation (NLSE). Using the perturbative method, we are going to approximate dispersion and nonlinearity to the lowest order. We introduce 'Peregrine' solution [15] as the lowest order treatment. However, more accurate models can be shown by including higher-order such as third order and fourth-order approximations. To obtain an analytic solution we apply similarity transformations [16]. We focused on the lowest order solution of NLSE with the linear potential which is time-dependent.

To transform KdV to NLSE we introduce a Fourier expansion of field variable as:

$$\psi = \sum_{m=1}^{\infty} \grave{o}^m \sum_{s=-m}^{m} \{\psi_s e^{is\Phi}\}, \psi_s = \sum_{s=0}^{\infty} \grave{o}^m \bar{\psi}_s^{(s)}$$

$\phi_0$ and $\psi_s$ are varying very slowly with time and space coordinates.

Expanding $\psi$ we get a form like this:

$$\psi = \left(\grave{o}\bar{\psi}_0 + \grave{o}\bar{\psi}_0^*\right) + \left(\grave{o}\bar{\psi}_0 e^{i\Phi} + \grave{o}\bar{\psi}_1^* e^{-i\Phi}\right) \\ + \left(\grave{o}^2\bar{\psi}_2 e^{2i\Phi} + \grave{o}^2\bar{\psi}_2^* e^{-2i\Phi}\right) + \ldots \quad (50)$$

Introducing a new stretching variable where 'c' is as group velocity,

$$\rho = \grave{o}\left[\xi - c_g \tau\right] \text{ and } \theta = \grave{o}^2 \tau$$

Using perturbation expression changing all variables in terms of $\rho$ and $\theta$ we get,

$$\frac{\partial}{\partial \tau} = -is\omega - \grave{o}c_g \frac{\partial}{\partial \rho} + \grave{o}^2 \frac{\partial}{\partial \theta}$$

$$\frac{\partial}{\partial \xi} = isk + \grave{o}\frac{\partial}{\partial \rho}$$

Now equating the coefficients of $e^{i\Phi}$ with $\grave{o}$ from perturbation expansion (14) we get,

$$\omega = -D_1 k^3 \quad (51)$$

$$c_g = \frac{\partial \omega}{\partial k} = -3D_1 k^2 \quad (52)$$

Again, equating the coefficient of $e^{2i\Phi}$ with $\grave{o}^2$ from perturbation expansion (14) we get,

$$\bar{\psi}_2^{(1)} = \frac{N_1}{6D_1 k^2} \bar{\psi}_1^{2(1)} \quad (53)$$

Again, equating the coefficients independent of $\Phi$ with $\grave{o}^3$ from perturbation expansion (14) we get,

$$\bar{\psi}_0^{(1)} = -\frac{C_1}{c_g} \bar{\psi}_1^{(1)} \bar{\psi}_1^{*(1)} \quad (54)$$

Where, $c_g$ is the group velocity of wave and N is the nonlinear term and D is the dispersive term in form of (28), (29). Equating for first harmonics,

$$\frac{\partial \bar{\psi}_1^{(1)}}{\partial \theta} + 3iD_1 k \frac{\partial^2 \bar{\psi}_1^{(1)}}{\partial \rho^2} - \frac{iC_1^2}{6D_1 k}\left(\bar{\psi}_1^{2(1)}\bar{\psi}_1^{*(1)}\right) = 0$$

$$i\frac{\partial \bar{\psi}_1^{(1)}}{\partial \theta} - 3D_1 k \frac{\partial^2 \bar{\psi}_1^{(1)}}{\partial \rho^2} + \frac{C_1^2}{6kD_1}\bar{\psi}_1^{2(1)}\bar{\psi}_1^{*(1)} = 0 \quad (55)$$

$$i\frac{\partial \bar{\psi}_1^{(1)}}{\partial \theta} + \frac{P}{2}\frac{\partial^2 \bar{\psi}_1^{(1)}}{\partial \rho^2} + Q\bar{\psi}_1^{2(1)}\bar{\psi}_1^{*(1)} = 0 \quad (56)$$

Again, considering $\bar{\psi}_1^{(1)} \equiv \Psi$.

$$i\frac{\partial \Psi}{\partial \theta} + \frac{P}{2}\frac{\partial^2 \Psi}{\partial \rho^2} + Q\Psi^2\Psi^* = 0 \quad (57)$$

Where, $P = -6D_1 k$ and $Q = \frac{C_1^2}{6kD_1} - N_1 k$. We see that $PQ < 0$ always, so the wave is always stable. Equation (57) is desired NLSE. $PQ < 0$ suggests that wave is creating a '*Bright–type*' envelope solitons [17].

Approaching to a *Peregrine* 'soliton' which reads to:

$$\Psi = \sqrt{\frac{P}{Q}}\left[\frac{4(1+2iP\theta)}{\left(1+4\rho^2+4P^2\theta^2\right)} - 1\right]e^{iP\theta} \quad (58)$$

$$|\Psi| = \left|\sqrt{\frac{P}{Q}}\right| \times \sqrt{\left(\frac{4}{\left(1+4\rho^2+4P^2\theta^2\right)} - 1\right)^2 + \frac{64P^2\theta^2}{\left(1+4\rho^2+4P^2\theta^2\right)^2}} \quad (59)$$

We see that the solution is $\rho$ and $\theta$ or $\xi$ and $\tau$ dependent thus waveform is nontrivial over small regions of $\xi$ and $\tau$. It is clear that $P < 0 < Q$
$PQ = -C_1^2 + 6N_1 D_1 k^2$, so plotting $PQ$ over different quantities like kappa $(\kappa)$, Mach number $(M)$, streaming velocity $(v_0)$ we can analyse stability properties of Rogue wave.

## 4. Dynamics of the system

After deriving the existence of peregrine profile this paper would determine how this wave propagates as well as how it evolves and its response to small perturbations. Differences in phase trajectory for systems exist with slightly different initial conditions. Similar studies on electron-acoustic super non-linear waves exist [18], [19] and also in Thomas-fermi plasma [20].

### 4.1 Unperturbed System

Using the coordinate transformation, $\chi = l\rho - V\theta$ and complex transformations, $\psi(\chi) = \tau(\chi)e^{i\beta\chi}$ on the equation (57) we obtain,

$$\left(\beta V \tau(\chi) - \frac{\rho l^2 \beta^2}{2}\tau(\chi) + \frac{\rho l^2}{2}\tau''(\chi) + Q\tau^3(\chi)\right) \\ + i\left(-V\tau'(\chi) + \rho l^2 \beta \tau'(\chi)\right)e^{i\beta\chi} = 0 \quad (60)$$

Comparing the real part on both sides we obtain,

$$\beta V \tau - \frac{\rho l^2 \beta^2}{2}\tau + \frac{\rho l^2}{2}\tau''(\chi) + Q\tau^3 = 0 \quad (61)$$

$$\Rightarrow \tau'' = \frac{\left(\frac{\rho l^2 \beta^2}{2} - \beta V\right)\tau}{\rho l^2 / 2} - \frac{Q\tau^3}{\rho l^2 / 2} \quad (62)$$

Where, $C = \frac{\left(\frac{\rho l^2 \beta^2}{2} - \beta V\right)\tau}{\rho l^2 / 2}$ and $D = \frac{Q}{\rho l^2 / 2}$

$$y = \frac{\partial \tau}{\partial \chi} \quad (64)$$

$$\Rightarrow y' = C\tau - D\tau^3 \quad (65)$$

### 4.2  Perturbed System

In above cases we worked with a particular system, has no completely left on its own. Now we shall investigate the evolution of a small periodic external driving force on the system. We choose the periodic force to be $f\cos\omega\tau$. We get a modified equation (65) as,

$$y' = C\tau - D\tau^3 + f\cos(\omega\tau) \quad (66)$$

Along with equation (65) it forms another set of ordinary differential equation.

### 5.  Results and Discussions

One-dimensional hydrodynamic model and reductive perturbation technique is used to study both linear dispersive effect of an unmagnetized plasma in Ion Acoustic wave. Dispersion relation is obtained in general case which dependents on $\kappa$.

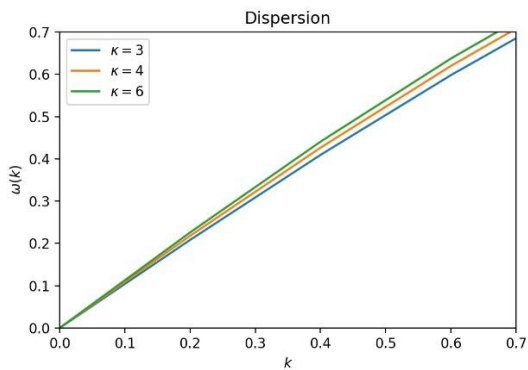

**Fig.2:** Dispersion curve between $\omega(k)$ vs $k$ for different $\kappa$ values.

We used $\mu = 1$ (electron-positron pair plasma) and $u_0 = 0.5$. From Fig. 2 we see that as κ value increases the frequency becomes increases; result is quite similar to [21]. So, this dispersive curve shows that there exists a propagation of linear wave in Ion Acoustic mode as we have considered hot electrons.

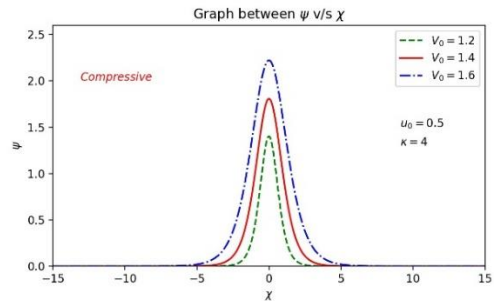

**Fig.3:** KdV solitary wave solution for Ion Acoustic mode $u_0 = 0.5, \kappa = 4, V_0 = 1.2, 1.4, 1.6$. *Bump-type solution*

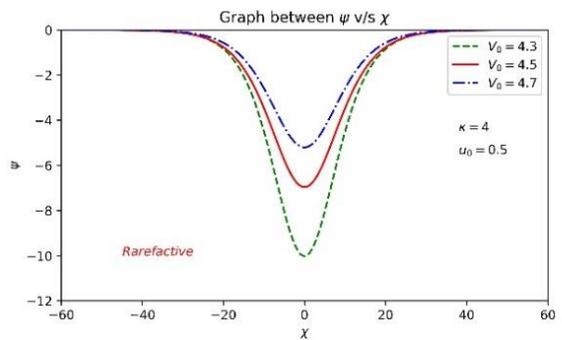

**Fig.4:** KdV solitary wave solution for Ion Acoustic mode $u_0 = 0.5, \kappa = 4, V_0 = 4.3, 4.5, 4.7$. *Dip-type solution*

To study nonlinear characteristics, we derived KdV equation with non-linearity and dispersive terms. KdV equation describes the solitary waves with small amplitude. We have studied the dependence of plasma parameters like streaming velocity $(u_0)$, Mach number $(M)$, mass ratio $(\mu), \kappa$ in details.

So, we can point out that solution (30) which can be expressed as Bump-type as well as Dip-type which are for Compressive and Rarefactive respectively. In Fig. 3 we have plotted potential with different wave velocities for a fixed $\kappa = 4$ value. For instance, when we take $\kappa = 4$ there exist both Bump and Dip-type solutions. The amplitude of the Bump-type solution increases with increasing of $V_0$. But as we increase further to the values $V_0 = 4.3$ the wave becomes Dip-type i.e., soliton becomes rarefactive. So, we can argue that there exists a critical value of $V_0$ between proposed values for that this certain transition occurs.

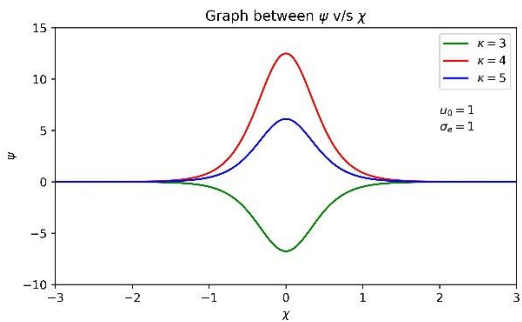

**Fig.5:** KdV solution for Ion Acoustic mode *(both Bump-type and Dip-type)* $u_0 = 1.0, \sigma_e = 1, \kappa = 3, 4, 5$

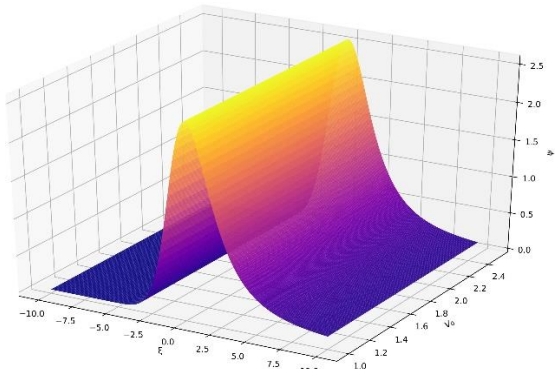

**Fig.6:** KdV solitary wave 3-dimensional simulation for compressive solution in Ion Acoustic mode $u_0 = 0.5, \kappa = 4$

Now we will observe the behaviour of the wave with different κ values in Fig. 5. So, we have plotted for $\kappa = 3.0, 4.0, 5.0$ fixed $V_0$ value. Between $\kappa = 4$ and $\kappa = 5$ amplitude decreases but as $\kappa$ decreases to 4 to 3 the amplitude becomes Bump-type type to Dip-type. This certain transition indicates that there exists a critical value of $\kappa$ between 3 and 4. The results are quite relatable with C.-R. Choi paper [21]. Now we have studied the first-order solution for Rogue wave and plotted it against $\theta$ and $\rho$ also projected along three planes. We adopted *ad hoc* values i.e., $P = Q = 1$. In this soliton excitation period is to increase with time as well as space. The spatial and temporal nature of RW is clearly analysed here. The peak determines the intense of potential. For fixed $\rho$ value there exists a critical $\theta(=\theta_c)$ value as well as for a fixed $\theta$ we get a critical $\rho(=\rho_c)$. By implying perturbating method, we evaluated the wave envelope which is shown in this Fig. 7.

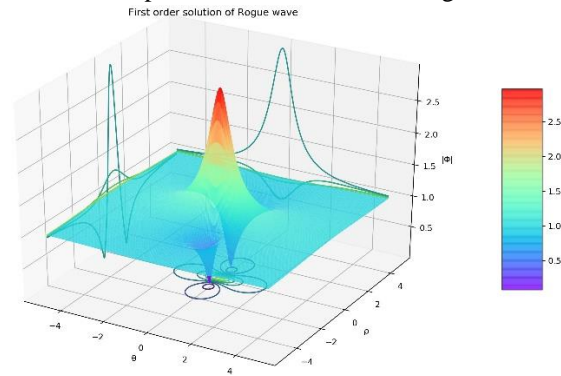

**Fig.7:** 'Peregrine' soliton of Rogue wave plotted against $\theta$ and $\rho$, have taken $P = Q = 1$

Clearly if we do $N_1 \to 0$ in Fig. 8 then *PQ* becomes negative so instability vanishes as well as the equation drops to KdV-NLSE. So, for KdV it always shows stable for that Rogue wave cannot generate.

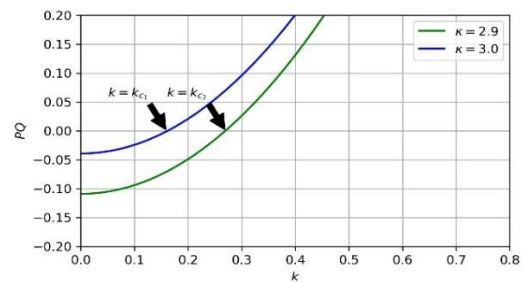

**Fig.8:** *PQ* plotted against *k* to find the stability, have taken $R = 1, \mu = 1$.

As we have plotted $PQ$ over $k$ to find the critical value $k$ for which it jumps from stable to instable mode. We conclude the critical case as for $Q = 0$. Where,

$$k = k_c = \left| \frac{C_1}{6D_1 N_1} \right|$$

Ion-acoustic Rogue wave is seen at laboratory [22] as well as stellar objects like Earth's atmosphere [23], Saturn's magnetosphere [24], and Solar wind plasma [25]. Equation (64) and (65) are the set of ordinary differential equations that are needed to be solved numerically in ordered to get the phase space plot of the rogue wave.

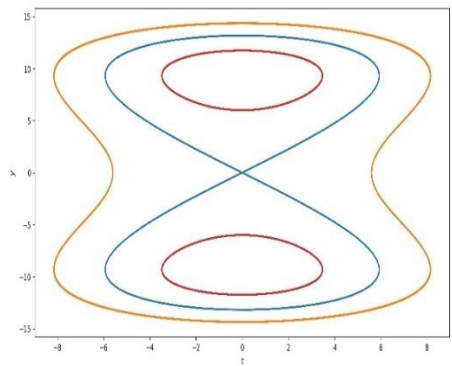

**Fig.9:** Phase trajectory of the system with initial conditions $(\tau_0, y_0) = (0.05, 0.05); (\tau_0, y_0) = (8,8);$ $(\tau_0, y_0) = (6, 0.05); (\tau_0, y_0) = (-6, 0.05)$ For orange, blue and red curves respectively. It is a phase trajectory that is the variation of the quasi velocity with quasi position, as quasi-time evolves.

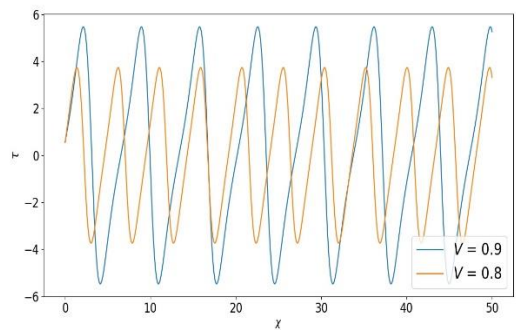

**Fig.10:** Time series for systems with same initial conditions, $(\tau_0, y_0) = (2.7, 0.55)$ but V = 0.9, and V = 0.8 for blue and orange curves respectively.

This is (Fig. 10) the time-series of rogue wave. Here we find two possible points (attractors) around which the system orbits. We get the phase trajectory for the case when we apply a small periodic driving force on the system. We took the magnitude and frequency of the force to be small. The initial conditions of the two systems are taken ad hoc but very close to the two attractors we found in the previous treatment.

Again, we employed the numerical technique and obtained the phase trajectory of the perturbed dynamical system. Here we have provided two different ad hoc initial conditions very close to each other to get an idea about the chaos of the system.

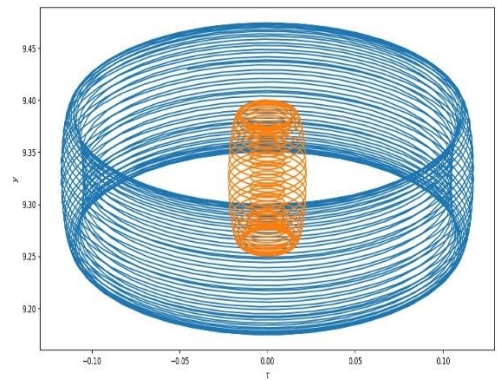

**Fig.11:** The amplitude and frequency of the driving force is given as, $f = 0.1$ and $\omega = 0.1$. Initial conditions are $(\tau_0, y_0) = (9.3, 0), (\tau_0, y_0) = (9.4, 0)$ for blue and orange curves respectively.

The difference in trajectories is in due to small differences in initial conditions. The sparking feature that reveals itself here is that there are further two attractors very close to each other. So, from our treatment we can say there are 4 attractors of the system.

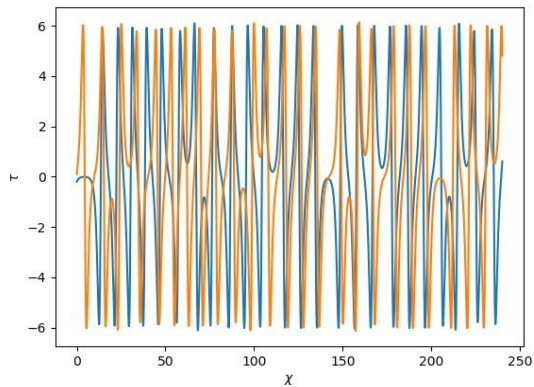

**Fig.12:** The time series for the systems with initial conditions $(\tau_0, y_0) = (0.1, -0.2)$, $(\tau_0, y_0) = (0.7, 0.1)$.

Without actually deriving the Lyapunov exponents, which is bit out of scope of this paper, we can informally remark that the system is chaotic with 4 attractors.

## 6. Conclusion

In summary we have discussed about the possibilities of generation of Rogue wave whether it could be generated for all case or not. We have noted that Rogue wave is experienced for a very short time period of time also it will generate for modified KdV version. For decrement of nonlinear quantity, the amplitude of wave envelops decreases. In NLSE the $Q$ factor is only one for which generation of Rogue wave can be decided. Also, we studied about critical values of different quantities. In this study we almost clear understanding of nonlinear excitation for Ion-acoustic mode. Phase trajectory of the Rogue wave has been studied for perturbed and isolated system, which has given us a brief insight into the dynamics of the system.

**Acknowledgements**

The authors would like to thank the reviewers for their valuable inputs towards upgrading the paper. We thank the Institute of Natural sciences and Applied Technology for providing facilities to carry out this work.